\documentclass[12pt,onecolumn,journal]{IEEEtran}

\usepackage{mathrsfs}
\usepackage{txfonts}
\usepackage [dvips] {graphicx}

\renewcommand\baselinestretch{2.0}

\ifCLASSINFOpdf

\else

\fi



\begin{document}

\title{Sidelobe Suppression for Capon Beamforming with Mainlobe to Sidelobe Power Ratio Maximization}

\author{Yipeng~Liu,~\
         and~Qun~Wan

\thanks{Yipeng Liu and Qun Wan are with the Electronic Engineering Department,
University of Electronic Science and Technology of China, Chengdu,
611731, China. Yipeng Liu is also a visiting researcher with
Electronic Engineering Department of the Electronic Engineering
Department, Tsinghua University, 100084, China.
e-mail: (liuyipeng@uestc.edu.cn; wanqun@uestc.edu.cn);.}
\thanks{Manuscript received Month Day, 2011; revised Month Day, Year.}}

\markboth{Journal Title,~Vol.~X, No.~X, Month~Year}%
{Shell \MakeLowercase{\textit{et al.}}: Bare Demo of IEEEtran.cls for Journals}

\maketitle

\begin{abstract}

High sidelobe level is a major disadvantage of the Capon
beamforming. To suppress the sidelobe, this paper introduces a
mainlobe to sidelobe power ratio constraint to the Capon
beamforming. it minimizes the sidelobe power while keeping the
mainlobe power constant. Simulations show that the obtained
beamformer outperforms the Capon beamformer.

\end{abstract}


\IEEEpeerreviewmaketitle

\section{Introduction}

\IEEEPARstart{A}{} beamformer is a multiple antennas system which
makes spatial beam focused on the target direction and spatial beam
nulled interference signal. In this system, each transmit antenna is
weighted by a properly designed gain and phase shift before
transmission. It is used extensively to improve the performance in a
variety of areas such as radar, sonar and wireless communications
\cite{li_rab}.

The Capon beamformer is one of the most popular beamforming systems.
It is a data-dependent beamformer which minimizes the array output
power subject to the linear constraint that the signal-of-interest
(SOI) does not suffer from any distortion by adaptive selection of
the weight vector. The Capon beamformer has better resolution and
much better interference rejection capability than the
data-independent beamformer. However, its high sidelobe level and
the SOI steering vector uncertainty due to differences between the
assumed signal arrival angle and the true arrival angle would
seriously degenerate the performance in the presence of environment
noise and interferences \cite{li_rab, wax_mvb, cox_mismatch}.

With a spherical uncertainty set was introduced, doubly constrained
robust (DCR) Capon beamformer is obtained with the increased
robustness against DOA mismatch \cite{li_dcrcb}. To achieve a faster
convergence speed and a higher steady state signal to interference
plus noise ratio (SINR), \cite{zhang_mvb_lqc} constrains its weight
vector to a specific conjugate symmetric form. In \cite{lie_ircb},
by iteratively estimation of the actual steering vector based on
conventional RCB formulation, iterative robust capon beamformer
(IRCB) with adaptive uncertainty level is proposed to enhance the
robustness against DOA mismatch.

Many works have been done to enhance the robustness against the SOI
steering vector uncertainty, to accelerate the convergence, etc
\cite{du_prrab}. However, no constraint is put onto the interference
and background noise. Based on the Capon criterion, a new constraint
is added to maximize the mainlobe and sidelobe power ratio (MSPR) to
shape the beam pattern. As more power is accumulated in the mianlobe
area, the robustness against SOI steering vector uncertainty is
improved; and as less power is in the sidelobe area, the
interference and noise rejection capability can be enhanced.
Therefore, a better performance can be achieved. Numerical Results
also demonstrate that the proposed beamformer outperforms.

\section{Signal Model}

The signal impinging into a uniform linear array (ULA) with \emph{M}
antennas can be represented by an \emph{M}-by-1 vector
\cite{li_rab}:
\begin{equation}
\label{eq2_1_received_signal} {\bf{x}}(k) = s(k){\bf{a}}({\theta
_0}) + \sum\limits_{j = 1}^J {{\beta _j}(k){\bf{a}}({\theta _j})}  +
{\bf{n}}(k)
\end{equation}
where \emph{k} is the index of time, \emph{J} is the number of
interference sources, \emph{s}(\emph{k}) and  \emph{j}(\emph{k})
(for \emph{j} = 1, ... , \emph{J}) are the amplitudes of the signal
of interest (SOI) and interfering signals at time instant \emph{k},
respectively,${{\theta _l}}$ (for \emph{l} = 0, 1, ... , \emph{J})
 are the direction of arrivals (DOAs) of the SOI and
interfering signals, $ {\bf{a}}({\theta _l}) = {\left[
{\begin{array}{*{20}{c}}
   1 & {\exp (j{\varphi _l})} &  \cdots  & {\exp (j(M - 1){\varphi _l})}  \\
\end{array}} \right]^T} $ (for \emph{l} = 0, 1, ... , \emph{J}) are the steering
vectors of the SOI and interfering signals, wherein $ {\varphi _l} =
({{2\pi d} \mathord{\left/
 {\vphantom {{2\pi d} \lambda }} \right.
 \kern-\nulldelimiterspace} \lambda })\sin {\theta _l} $, with \emph{d}
being the distance between two adjacent sensors and $ \lambda $
being the wavelength of the SOI; and \textbf{n}(\emph{k}) is the
additive white Gaussian noise (AWGN) vector at time instant
\emph{k}.

The output of a beamformer for the time instant \emph{k} is then
given by:
\begin{equation}
\label{eq2_2_array_output}y(k) = {{\bf{w}}^H}{\bf{x}}(k) =
s(k){{\bf{w}}^H}{\bf{a}}({\theta _0}) + \sum\limits_{j = 1}^J
{{\beta _j}(k){{\bf{w}}^H}{\bf{a}}({\theta _j})}  +
{{\bf{w}}^H}{\bf{n}}(k)
\end{equation}
where \emph{\textbf{w}} is the \emph{M}-by-1complex-valued weighting
vector of the beamformer.

\section{The Proposed Beamformer}

The Capon beamformer is defined as the solution to the following
linearly constrained minimization problem \cite{li_rab}:
\begin{equation}
\label{eq3_1_Capon} {{\bf{w}}_{Capon}} = \mathop {\arg \min
}\limits_{\bf{w}} \left( {{{\bf{w}}^H}{{\bf{R}}_x}{\bf{w}}}
\right),{\rm{  s}}{\rm{.t}}{\rm{.  }}{{\bf{w}}^H}{\bf{a}}({\theta
_0}) = 1
\end{equation}
where $ {{{\bf{R}}_x}} $ is the \emph{M}-by-\emph{M} covariance
matrix of the received signal vector \textbf{x}(\emph{k}), and $
{{\bf{w}}^H}{\bf{a}}({\theta _0}) = 1 $ is the distortionless
constraint applied on the SOI.

In the perspective of the beam pattern, it is observed from the
Capon beeamformer (\ref{eq3_1_Capon}) that there is only an explicit
constraint on the desired DOA, i.e. $ {{\bf{w}}^H}{\bf{a}}({\theta
_0}) = 1 $, while no constraint is put onto the interference and
background noise. To repair this drawback, we propose the following
cost function with a regularization term, which forces maximization
of the MSPR:
\begin{equation}
\label{eq3_2_MSPR}\begin{array}{c}
 {{\bf{w}}_{MSPR}} = \mathop {\arg \min }\limits_{\bf{w}} \left\{ {{{\bf{w}}^H}{{\bf{R}}_x}{\bf{w}} + \gamma \left[ {{{\left( {\left\| {{{\bf{w}}^H}{{\bf{A}}_M}} \right\|_2^2 - 1} \right)}^2} + \left\| {{{\bf{w}}^H}{{\bf{A}}_S}} \right\|_2^2} \right]} \right\} \\
 {\rm{  s}}{\rm{.t}}{\rm{.  }}{{\bf{w}}^H}{\bf{a}}({\theta _0}) = 1 \\
 \end{array}
\end{equation}
where $ \gamma $ is the weighting factor balancing the minimum
variance constraint and the MSPR minimization constraint. The
\emph{M}-by-\emph{N} \textbf{A} is the array manifold with $ \alpha
$ns ( \emph{n} = 1, 2, ... , \emph{N} ) being the sampled angles in
the [ $ - {90^ \circ } $, $ {90^ \circ } $], and the \emph{N}
steering vectors cover all the DOAs in the sampling range, with $
{\alpha _0} $ being the DOA of the SOI as defined in
(\ref{eq2_1_received_signal}), i.e.,
\begin{equation}
\label{eq3_3_MSPR1} {A_{mn}} = \exp \left( {j(m - 1){\varphi _n}}
\right),{\rm{  for }}m = 1,...,M; n = 1,...,N{\rm{ }}
\end{equation}

\begin{equation}
\label{eq3_3_MSPR2}{\varphi _n} = \frac{{2\pi d}}{\lambda }\sin
{\alpha _n},{\rm{ for }}n = 1,...,N
\end{equation}

\begin{equation}
\label{eq3_4_A_M}{{\bf{A}}_M} = \left[ {\begin{array}{*{20}{c}}
   {{\bf{a}}\left( {{\theta _{ - b}}} \right)} &  \cdots  & {{\bf{a}}({\theta _0})} &  \cdots  & {{\bf{a}}({\theta _{ + b}})}  \\
\end{array}} \right]
\end{equation}

\begin{equation}
\label{eq3_5_A_S}{{\bf{A}}_S} = \left[ {\begin{array}{*{20}{c}}
   {{\bf{a}}({\theta _{ - 90}})} &  \cdots  & {{\bf{a}}({\theta _{ - b - 1}})} & {{\bf{a}}({\theta _{ + b + 1}})} &  \cdots  & {{\bf{a}}({\theta _{ + 90}})}  \\
\end{array}} \right]
\end{equation}

$ {{\bf{A}}_S} $ is sub-matrix of the steering matrix \textbf{A},
and it is constituted with the sidelobe steering vectors in
\textbf{A}. $ {{\bf{A}}_M} $ is sub-matrix of the steering matrix
\textbf{A} too, and it is constituted with the mainlobe steering
vectors in \textbf{A}. \emph{b} is an integer corresponding to the
bounds between the mainlobe and the sidelobe of the beam pattern.

The product $ {{\bf{w}}^H}{{\bf{A}}_S} $ indicates array gains of
the sidelobe; and the product $ {{\bf{w}}^H}{{\bf{A}}_M} $ indicates
array gains of the sidelobe. The newly added MSPR $ {\left( {\left\|
{{{\bf{w}}^H}{{\bf{A}}_M}} \right\|_2^2 - 1} \right)^2} + \left\|
{{{\bf{w}}^H}{{\bf{A}}_S}} \right\|_2^2 $ is minimized to minimize
the sidelobe power $ \left\| {{{\bf{w}}^H}{{\bf{A}}_S}} \right\|_2^2
$ while enforcing the mainlobe power to be a constant. Thus, the
MSPR $ {{\left\| {{{\bf{w}}^H}{{\bf{A}}_M}} \right\|_2^2}
\mathord{\left/
 {\vphantom {{\left\| {{{\bf{w}}^H}{{\bf{A}}_M}} \right\|_2^2} {\left\| {{{\bf{w}}^H}{{\bf{A}}_S}} \right\|_2^2}}} \right.
 \kern-\nulldelimiterspace} {\left\| {{{\bf{w}}^H}{{\bf{A}}_S}} \right\|_2^2}} $ would be maximized.

The proposed optimization model can be solved efficiently by
Lagrange multiplier method. First the Lagrange multipliers technique
is used to combine the constraints in (\ref{eq3_2_MSPR}) into the
objective function:

\begin{equation}
\label{eq3_6_Lagrange}\begin{array}{c}
 f({\bf{w}}) = {{\bf{w}}^H}{{\bf{R}}_x}{\bf{w}} + \gamma {\left( {\left\| {{{\bf{w}}^H}{{\bf{A}}_M}} \right\|_2^2 - 1} \right)^2} \\
  + \gamma \left\| {{{\bf{w}}^H}{{\bf{A}}_S}} \right\|_2^2 + \mu \left( {{{\bf{w}}^H}{\bf{a}}({\theta _0}) - 1} \right) \\
  = {{\bf{w}}^H}{{\bf{R}}_x}{\bf{w}} + \gamma {\left( {{{\bf{w}}^H}{{\bf{A}}_M}{\bf{A}}_M^H{\bf{w}} - 1} \right)^2} \\
  + \gamma {{\bf{w}}^H}{{\bf{A}}_S}{\bf{A}}_S^H{\bf{w}} + \mu \left( {{{\bf{w}}^H}{\bf{a}}({\theta _0}) - 1} \right) \\
 \end{array}
\end{equation}

Thus, we have:
\begin{equation}
\label{eq3_7_Lagrange1}\begin{array}{l}
 \frac{{\partial f({\bf{w}},\mu )}}{{\partial {{\bf{w}}^H}}} = {{\bf{R}}_x}{\bf{w}} + \gamma \left( {{{\bf{w}}^H}{{\bf{A}}_M}{\bf{A}}_M^H{\bf{w}} - 1} \right)\left( {2{{\bf{A}}_M}{\bf{A}}_M^H{\bf{w}}} \right) \\
 {\rm{~~~~~~~~~~}} + \gamma {{\bf{A}}_S}{\bf{A}}_S^H{\bf{w}} + \mu {\bf{a}}({\theta _0}) \\
 \end{array}
\end{equation}

\begin{equation}
\label{eq3_8_Lagrange2}\frac{{\partial f({\bf{w}},\mu )}}{{\partial
\mu }} = {{\bf{w}}^H}{\bf{a}}({\theta _0}) = 1
\end{equation}

From (\ref{eq3_7_Lagrange1}),

\begin{equation}
\label{eq3_9_Lagrange3}\left( {{{\bf{R}}_x} + \gamma \left(
{2{{\bf{w}}^H}{{\bf{A}}_M}{\bf{A}}_M^H{\bf{w}} - 2}
\right){{\bf{A}}_M}{\bf{A}}_M^H + \gamma {{\bf{A}}_s}{\bf{A}}_S^H}
\right){\bf{w}} =  - \mu {\bf{a}}({\theta _0})
\end{equation}

\begin{equation}
\label{eq3_10_Lagrange4}{\bf{w}} =  - \mu {\left( {{{\bf{R}}_x} +
\gamma \left( {2{{\bf{w}}^H}{{\bf{A}}_M}{\bf{A}}_M^H{\bf{w}} - 2}
\right){{\bf{A}}_M}{\bf{A}}_M^H + \gamma {{\bf{A}}_s}{\bf{A}}_S^H}
\right)^{ - 1}}{\bf{a}}({\theta _0})
\end{equation}

Substituting (\ref{eq3_10_Lagrange4}) into (\ref{eq3_8_Lagrange2})
gives:
\begin{equation}
\label{eq3_11_Lagrange5} - \mu {\bf{a}}{({\theta _0})^H}{\bf{\Xi
a}}({\theta _0}) = 1
\end{equation}
where
\begin{equation}
\label{eq3_12_Lagrange6}{\bf{\Xi }} = {\left( {{{\left(
{{{\bf{R}}_x} + \gamma \left(
{2{{\bf{w}}^H}{{\bf{A}}_M}{\bf{A}}_M^H{\bf{w}} - 2}
\right){{\bf{A}}_M}{\bf{A}}_M^H + \gamma {{\bf{A}}_s}{\bf{A}}_S^H}
\right)}^H}} \right)^{ - 1}}
\end{equation}

Then,
\begin{equation}
\label{eq3_13_Lagrange7} \mu  = \frac{{ - 1}}{{{\bf{a}}{{({\theta
_0})}^H}{\bf{\Xi a}}({\theta _0})}}
\end{equation}

Substituting (\ref{eq3_13_Lagrange7}) into (\ref{eq3_10_Lagrange4}),
gives:

\begin{equation}
\label{eq3_14_Lagrange8} {\bf{w}} = \frac{{{\bf{\Xi a}}({\theta
_0})}}{{{\bf{a}}{{({\theta _0})}^H}{\bf{\Xi a}}({\theta _0})}}
\end{equation}

Therefore, the iterative algorithm is:

\begin{equation}
\label{eq3_15_Lagrange9} \begin{array}{c}
 {\bf{w}}(i + 1) \\
  = \frac{{{{\left( {{{\bf{R}}_x} + \gamma \left( {2{\bf{w}}{{(i)}^H}{{\bf{A}}_M}{\bf{A}}_M^H{\bf{w}}(i) - 2} \right){{\bf{A}}_M}{\bf{A}}_M^H + \gamma {{\bf{A}}_s}{\bf{A}}_S^H} \right)}^{ - 1}}{\bf{a}}({\theta _0})}}{{{\bf{a}}{{({\theta _0})}^H}{{\left( {{{\left( {{{\bf{R}}_x} + \gamma \left( {2{\bf{w}}{{(i)}^H}{{\bf{A}}_M}{\bf{A}}_M^H{\bf{w}}(i) - 2} \right){{\bf{A}}_M}{\bf{A}}_M^H + \gamma {{\bf{A}}_s}{\bf{A}}_S^H} \right)}^H}} \right)}^{ - 1}}{\bf{a}}({\theta _0})}} \\
 \end{array}
\end{equation}
where \emph{i} denotes the interaction index.

\section{Simulation}

In the simulations, a ULA with 8 half-wavelength spaced antennas is
considered. The AWGN at each sensor is assumed spatially
uncorrelated. The DOA of the SOI is set to be $ {0^ \circ } $, and
the DOAs of three interfering signals are set to be  $  - {30^ \circ
} $, $  {30^ \circ } $, and $  {70^ \circ } $, respectively. The
signal to noise ratio (SNR) is set to be 10 dB, and the interference
to noise ratios (INRs) are assumed to be 20 dB, 20 dB, and 40 dB in
$  - {30^ \circ } $, $  {30^ \circ } $, and $  {70^ \circ } $,
respectively. 100 snapshots are used for each simulation. Without
loss of generality, b is set to be 12, $\gamma $  is chosen to be 1.
The matrix \textbf{A} consists of all steering vectors in the DOA
range of [$  - {90^ \circ } $, $  {90^ \circ } $] with the sampling
interval of $  {1^ \circ } $.

To examine its influence on the performance, the SINR is calculated
via the following formula:
\begin{equation}
\label{eq4_1_SINR} SINR = \frac{{\sigma
_s^2{{\bf{w}}^H}{\bf{a}}({\theta _0}){{\bf{a}}^H}({\theta
_0}){\bf{w}}}}{{{{\bf{w}}^H}\left( {\sum\limits_{j = 1}^J {\sigma
_j^2{\bf{a}}({\theta _j}){{\bf{a}}^H}({\theta _j})}  + {\bf{Q}}}
\right){\bf{w}}}}
\end{equation}
where $ \sigma _s^2 $ and $ \sigma _j^2 $  are the variances of the
SOI and \emph{j}-th interference, \textbf{Q} is a diagonal matrix
with the diagonal elements being the noise's variances.

\begin{figure}[!h]
 \centering
 \includegraphics[angle= 0, scale = 0.47]{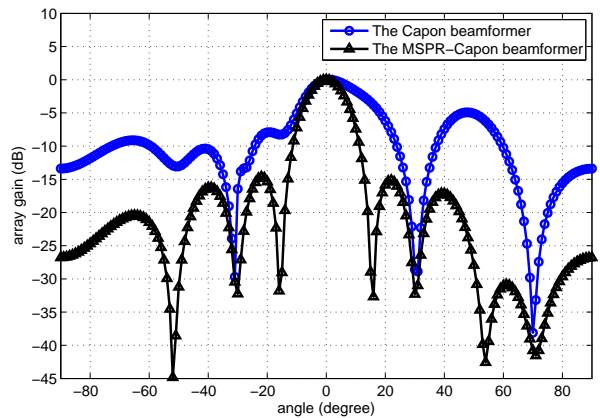}
 \caption{Normalized beam patterns of the Capon beamformer, and the MSPR-Capon beamformer, without mismatch between the steering angle and the DOA of the SOI.}
 \label{fig:1}
\end{figure}

Fig. \ref{fig:1} shows beam patterns of the Capon beamformer
(\ref{eq3_1_Capon}), the MSPR-Capon beamformer (\ref{eq3_2_MSPR}) of
1000 Monte Carlo simulations. It is obvious that a better sidelobe
suppression performance is achieved by the MSPR-Capon beamformer.
The beam pattern of the MSPR-Capon beamformer has a lower array gain
level in sidelobe area, and provides deeper nulls in the directions
of interference, i.e., $  - {30^ \circ } $, $  {30^ \circ } $, and $
{70^ \circ } $, respectively. The average received SINR by the Capon
beamformer, the MSPR-Capon beamformer (\ref{eq3_2_MSPR}) are 3.6809
dB, and 6.5224 dB. Besides when we change the parameter , in 1000
Monte Carlo simulations, the influence of choice of the parameter on
the output SINR is slight, and all the values of SINR lies between
6.4287dB and 6.7861dB.

\begin{figure}[!h]
 \centering
 \includegraphics[scale = 0.47]{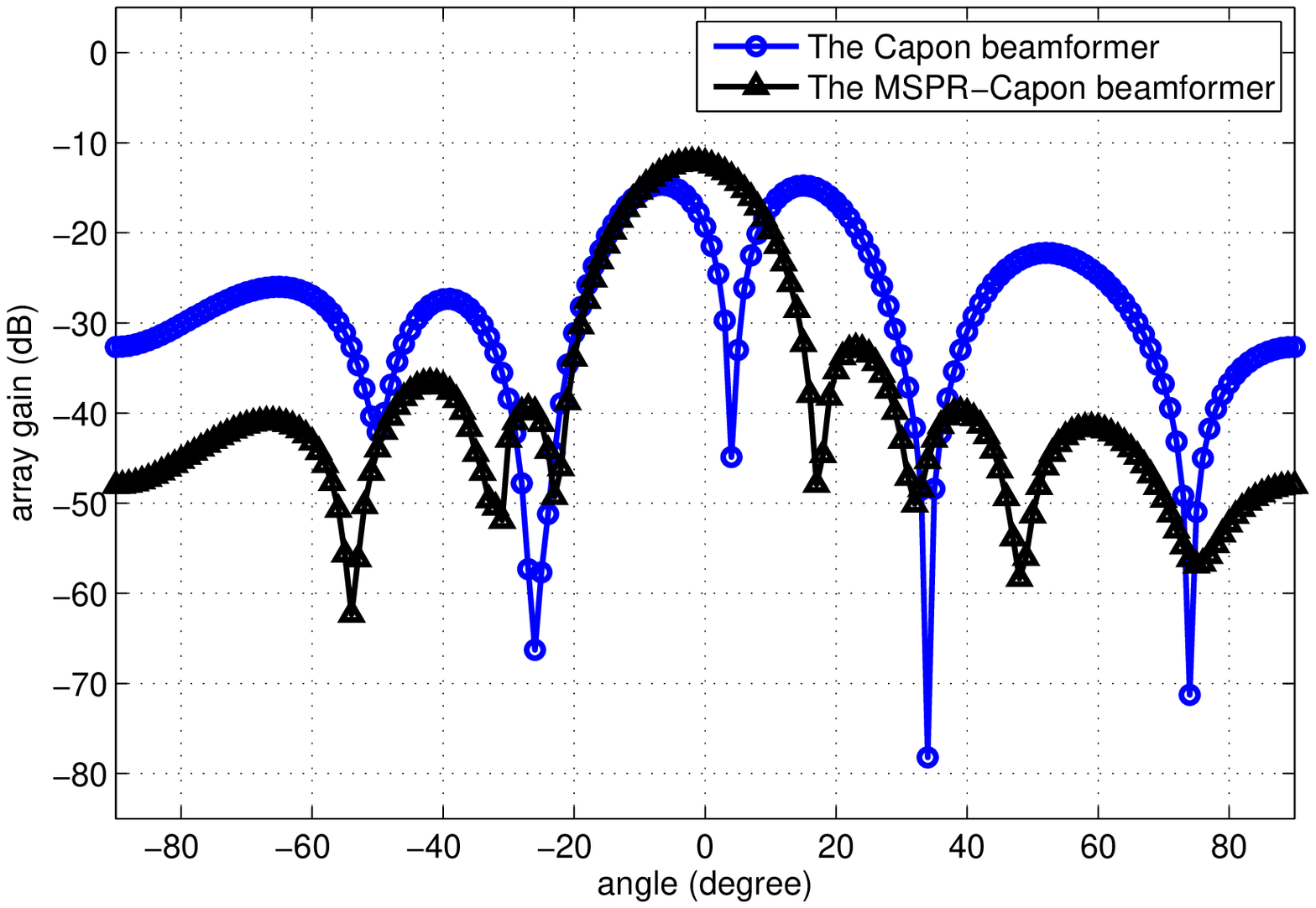}
 \caption{Normalized beam patterns of the Capon beamformer and the MSPR-Capon beamformer, with $ {4^ \circ } $ mismatch between the steering angle and the DOA of the SOI.}
 \label{fig:2}
\end{figure}

Fig. \ref{fig:2} shows beam patterns of the beamformers that we have
discussed, with each beamformer having a $  {4^ \circ } $ mismatch
between the steering angle and the DOA of the SOI. We can see that
the Capon beamformer has a deep notch in $  {4^ \circ } $, which is
the DOA of the SOI. It can be explained by using the fact that the
Capon beamformer is designed to minimize the total array output
energy subject to a distortionless constraint in the DOA of the SOI,
so when the steering angle is in $  {4^ \circ } $, instead of $ {0^
\circ } $, the Capon beam pattern maintains distortionless in $ {0^
\circ } $ while resulting in a deep null in $ {4^ \circ } $. This
observation shows the high sensitivity of the Capon beamformer to
steering angle mismatch. Comparing beam patterns of beamformers
defined in (\ref{eq3_1_Capon}) and (\ref{eq3_2_MSPR}), we can see
that the MSPR-Capon (\ref{eq3_2_MSPR}) has a high array gain in the
DOA of the SOI. Besides, it suppresses sidelobe levels and deepens
the nulls for interference avoidance. In the case of $ {4^ \circ } $
mismatch, the average received SINR by the Capon beamformer
(\ref{eq3_1_Capon}) and the MSPR-Capon (\ref{eq3_2_MSPR}) are 0.0009
dB and 3.8402 dB respectively. 1000 Monte Carlo simulations show
that the influence of choice of the parameter on the output SINR is
slight too, and all the values of SINR lie between 3.6571dB and
3.9380 dB.

\section{Conclusion}

The proposed beamformer introduces a new beam pattern shaping
constraint. it shows superiority to the MVDR beamformer. The
problems of the Capon beamformer's high sidelobe level as well as
sensitivity to SOI steering vector errors are much alleviated. In
the future work, the MSPR can be incorporated to other Capon based
beamformer, such as DCR Capon beamformer, to further enhance the
performance.

\section*{Acknowledgment}

This work was supported in part by the National Natural Science
Foundation of China under grant 60772146, the National High
Technology Research and Development Program of China (863 Program)
under grant 2008AA12Z306 and in part by Science Foundation of
Ministry of Education of China under grant 109139.

\ifCLASSOPTIONcaptionsoff
  \newpage
\fi

\end{document}